# 3D nano-bridge-based SQUID susceptometers for scanning magnetic imaging of quantum materials

Y. P. Pan[1,2], S. Y. Wang[1], X. Y. Liu[2], Y. S. Lin[1], L. X. Ma[2], Y. Feng[1], Z. Wang[2], L. Chen[2,*], Y. H. Wang[1,*]

1. Department of Physics and State Key Laboratory of Surface Physics, Fudan University, Shanghai 200438 China

2. Center for Excellence in Superconducting Electronics, State Key Laboratory of Functional Material for Informatics, Shanghai Institute of Microsystem and Information Technology, Chinese Academy of Sciences, Shanghai 200050 China

* To whom correspondence and requests for materials should be addressed.

Email: leichen@mail.sim.ac.cn ; wangyhv@fudan.edu.cn

**Abstract**

**We designed and fabricated a new type of superconducting quantum interference device (SQUID) susceptometers for magnetic imaging of quantum materials. The 2-junction SQUID sensors employ 3D Nb nano-bridges fabricated using electron beam lithography. The two counter-wound balanced pickup loops of the SQUID enable gradiometric measurement and they are surrounded by a one-turn field coil for susceptibility measurements. The smallest pickup loop of the SQUIDs were 1 μm in diameter and the flux noise was around 1 $\mu\Phi_0/\sqrt{Hz}$ at 100 Hz. We demonstrate scanning magnetometry, susceptometry and current magnetometry on some test samples using these nano-SQUIDs.**

# Introduction

The superconducting quantum interference devices (SQUIDs)[1] are one of the most sensitive magnetic flux detectors available. Scanning SQUID microscopy (sSQUID)[2] uses a nano-fabricated SQUID (nano-SQUID)[3][4][5] as a local magnetometer to form magnetic images by scanning over the sample[6]. It collects flux through its pickup loop[7] due to DC magnetic fields as well as magnetic response of micro and nanometer scale structures, revealing electro-magnetic properties of materials that cannot be probed directly by other methods. sSQUID played an important role in determining the pairing symmetry in cuprate and iron-based high temperature superconductors[8]. Recently, it has been applied successfully to image the edge currents in HgTe/CdTe and InAs/GaSb heterostructures[9][10].

Many nano-SQUIDs[11] have been developed in recent years for various applications. For instance, Pb SQUID-on-tip[12][13][14] technology have achieved extremely high resolution and spin sensitivity. Nano-SQUIDs based on Nb/AlOx/Nb tri-layer junctions have also been developed to perform magnetic and susceptibility imaging simultaneously[15][16]. Planar Nb Dayem-bridge junctions were used in nano-SQUIDs fabricated by FIB to image the vectorial current density in 2D electron gas[17]. Dayem-bridge junctions[18][19] bear a much higher working field than that of the tri-layer junctions because of its much reduced area. Because a large magnetic field is an important tuning parameter in condensed matter systems, batch fabricated nano-SQUIDs with Dayem-bridge junctions are desirable as ultra-sensitive magnetometers for the study of quantum materials. However, planar Dayem-bridges suffer from parasitic capacitance and relatively large flux noise, both of which compromise the sensitivity of a magnetometer.

Here, we report the design, fabrication and characterization of nano-SQUIDs based on 3D nano-bridge junctions[20] for magnetic imaging of quantum materials. The non-planar geometry of the 3D nano-bridge significantly reduces parasitic capacitance[21] and flux noise[22][23]. Our nano-SQUID is designed as a gradiometer so that it is not sensitive to a uniformly applied magnetic field[16], which is suitable for scanning magnetometry under external field. Field coils in our nano-SQUIDs allow local measurement of susceptibility, which is very important for the study of 2D superconductors [monolayer FeSe]. Our nano-SQUIDs showed a working field range up to 5 KGs. The flux noise at 4.2 K was around 1 $\mu\Phi_0/Hz^{1/2}$ at 100 Hz under zero applied field, comparable to that of SQUIDs based on tri-layer junctions. We were able to demonstrate highly sensitive scanning magnetometry, susceptometry and current magnetometry using these nano-SQUIDs.

# Design and fabrication

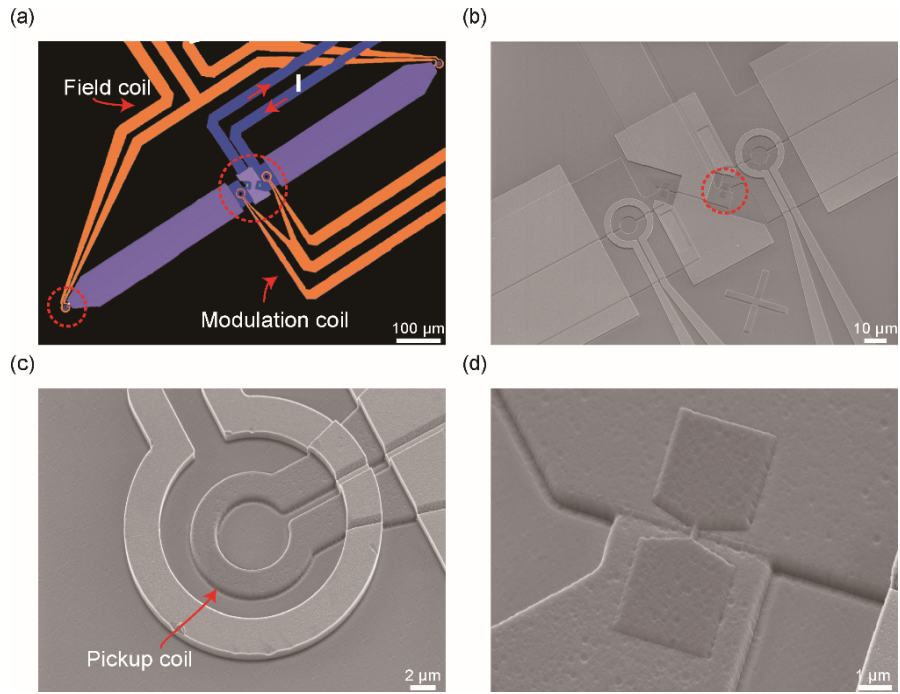

**Fig. 1. Layout and SEM images of the nano-SQUIDs.** (a) is a layout of our device. Pickup coils, field coils and modulation coils are circled with red dashed lines. (b) is an SEM image of the middle dotted circle in (a). (c) is an zoom-in view of the front field coil. (d) is a zoom-in view of the nano-bridge as circled in (b). Its width is 50 nm.

The layout of the nano-SQUID susceptometer consists of a SQUID loop with pickup coils, field coils and modulation coils (Fig. 1a). In order to reduce the inductance of the SQUID loop, it has a very narrow width and is covered with shielding except for the area of the pickup coils and modulations coils. The pickup coils are counter-wound loops at two ends of the SQUID loop to cancel any environmental uniform background signal. They are separated by 1 mm (or 0.5 mm in a later design) from each other. This way, one of them can be positioned within 1 μm from the sample while the other one is far away so that the magnetic signal from the sample only goes through the front pickup coil. The modulation coils are essential for flux-locked-loop (FLL) operation and thus direct readout of the flux signal. The field coils are used to apply a local perturbing magnetic field to the sample so that its response can be detected in the pickup coils as a susceptibility signal. The field coils are not counter-wound (Fig. 1a) so that the mutual inductance between the field coil and the pickup coils are minimized. The spatial separation of these coils from the junctions reduces their cross coupling and allows signal flux to more efficiently couple into the pickup coil[24].

The two junctions of our nano-SQUIDs were made of 3D Nb nano-bridges (Fig.1d) fabricated by electron-beam lithography (EBL). The advantage of nano-bridges over tri-layer junctions is that shunt resistance necessary for non-hysteretic operation of a SQUID magnetometer is inherent to the bridges. Since shunt resistors have to be made from non-superconducting metallic layers, their electrical contact with the Nb layers were generally difficult to make. Eliminating the step for shunt resistance greatly simplified our fabrication process of nano-SQUIDs (Fig. 2). Nevertheless, the design requirement of the counter-wound pickup coils, field coils and modulation coils dictates a multi-metal-layer fabrication process which we detail below.

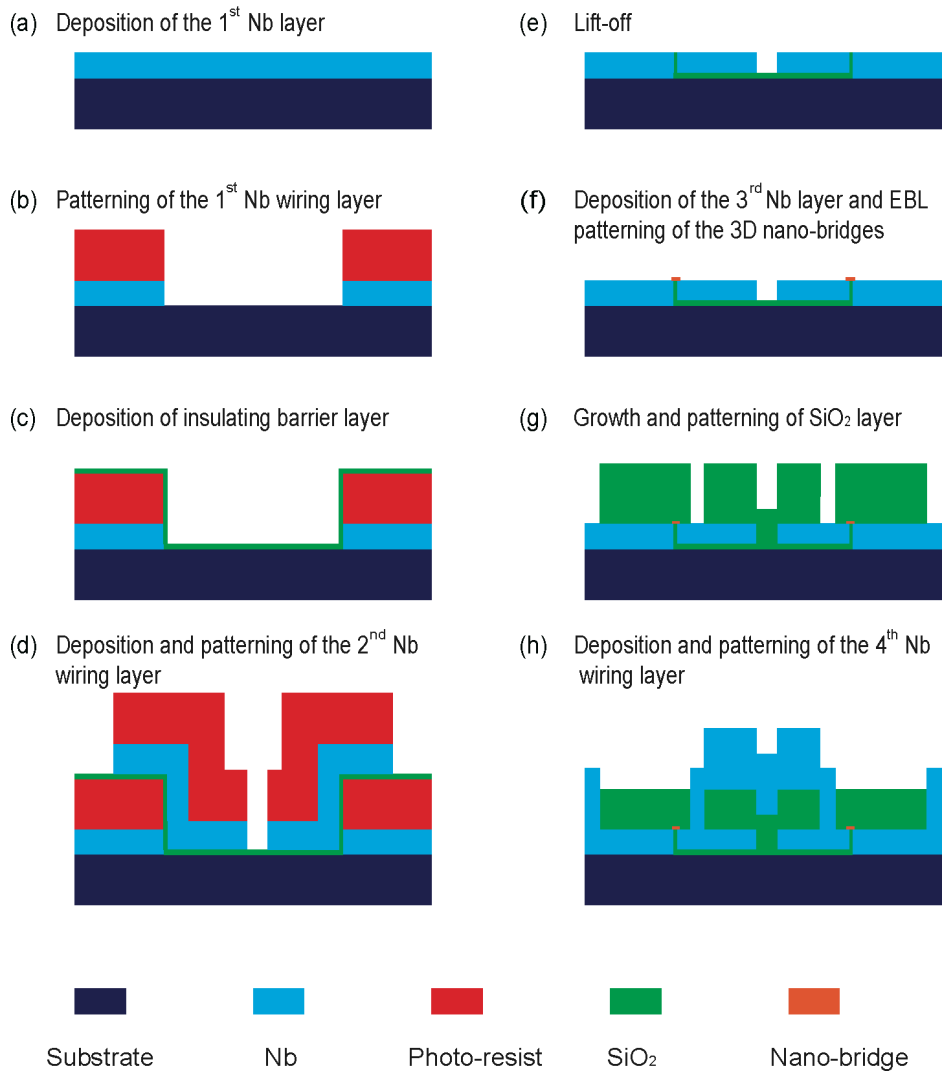

**Fig. 2 . Fabrication process of the 3D nano-bridge SQUIDs.** (a) Deposition of the 1st Nb layer (150 nm) on Si substrate with 300nm $SiO_2$ coating. (b) Photolithography and RIE to pattern the 1st Nb wiring layer. The photo-resist was not removed after etching in this step. (c) Deposition of 25 nm insulating barrier ($SiO_2$ or MgO) to cover the resist. (d) Deposition of the 2nd Nb layer (200 nm) and a subsequent photolithography and RIE to pattern it. (e) Lift-off that removed all previous photo-resist and exposed the grooves for the junctions. (f) Deposition of the 3rd Nb layer and patterning of the nano-bridges by EBL. (g) Growth of 250 nm of $SiO_2$ by PECVD and photolithography and RIE on the dielectric to make via's to the 2$^{nd}$ Nb layer. (h) Deposition of the 4$^{th}$ Nb (300 nm) followed by the last photolithography and RIE patterning.

The nano-SQUIDs fabrication process was developed based on 3D nano-bridge junctions. The major steps of fabrication process included four Nb wiring layers separated by SiO$_2$ as the dielectric, four photolithography steps and one EBL step (Fig. 2). The Nb wiring layers were deposited by magnetron sputtering. First, we grew a 150 nm Nb layer on a 4" Si wafer with 300 nm SiO$_2$ coating. We patterned the first layer by photolithography and reactive-ion etching machine (RIE) (Fig. 2a,b). The first layer contained the bias electrodes for the SQUID loop. The photo-resist (AZ MiR 703) was kept on in this step which we used later to make the 3D junction structure. 25 nm of SiO$_2$ (or MgO) and 200 nm of Nb were subsequently deposited on top of the photo-resist (Fig. 2c,d). The 2$^{nd}$ Nb layer was designed to be thicker to compensate the shadow effect of deposition near the edge of the 1$^{st}$ Nb layer. Another round of photolithography and RIE were performed to pattern the second Nb wiring layer. This layer and the insulating groove made up the two arms of the SQUID loop. Then we removed the photo-resist from the first two rounds of photolithography together by immersing the whole wafer into heated acetone for several hours (Fig. 2e). We found that hotter acetone solvent could facilitate the lift-off procedure. After the photo-resist were gone, we deposited the 3$^{rd}$ Nb layer and patterned the two nano-bridge junctions by EBL and RIE (Fig. 2f). For our nano-SQUIDs, we found the optimal nano-bridges were 15 nm thick and 50 nm wide. The advantage of using EBL is that the thickness and width of the nano-bridge, which is critical to the performance of the device, could be changed as desired easily. Due to the gradiometer design, the SQUID loop needs another (4$^{th}$) layer of Nb (about 300 nm thick). We grew a 250 nm SiO$_2$ layer by plasma enhanced chemical vapor deposition (PECVD) at 80℃ (Fig. 2g) to separate them. Two Nb via's were made to connect the 2$^{nd}$ Nb wiring layer to the 4$^{th}$ layer (Fig. 2h) in order to complete the SQUID loop. Part of the 4$^{th}$ layer of Nb was also used as shielding for the SQUID loop. Other parts of the 4$^{th}$ wiring layer are used for the field coils and modulation coils. The wafer was fully patterned after the etching of the 4$^{th}$ Nb. Lastly, we diced up the wafer for electronic characterizations.

# Device characterization and scanning magnetometry imaging

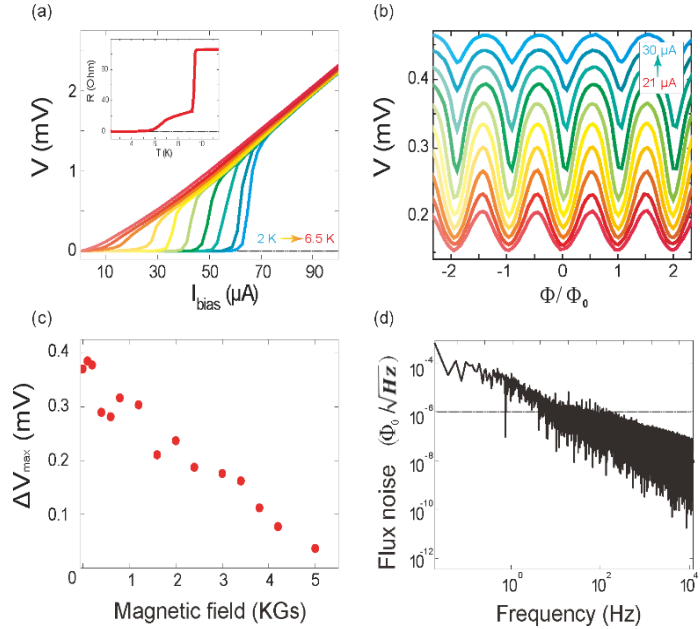

**Fig. 3. Electrical transport characterization of the nano-SQUIDs.** (a) Current-voltage characteristics of the device at different temperatures. Inset: resistance of the SQUID as a function of temperature. (b) SQUID voltage modulated by flux under different bias current at 5 K. The flux was applied through the modulation coils. (c) shows the modulation amplitude as a function of a perpendicular magnetic field. (d) Typical flux noise spectrum at 4.2 K measured under a flux-locked loop.

We carried out transport measurements of our nano-SQUIDs to characterize its noise performance (Fig. 3). A typical device showed two transition temperatures in its resistance (Fig. 3a inset). The higher one around 9.3 K was due to the Nb wirings making up the SQUID loop and the electrodes and the lower one around 6.5 K was from the Nb nano-bridges which were much thinner than the wiring layers. The Tc of our Nb film was very close to that of bulk Nb crystals[25], suggesting good quality of our films. Consistent with the resistance measurement, the current-voltage (IV) characteristics showed non-linear dependence below 6.5 K and reached around 60 uA of critical current[11] ($I_c$) at 2 K (Fig. 3a). There was no observable hysteresis in the IV's which is desirable for magnetometry.

As expected for a SQUID, the voltage across our devices were modulated by the flux through the SQUID loop (Fig. 3b). Such modulation could be applied through either the modulation coils or the field coils. The amplitude of the modulation was the largest when the bias was at Ic. This amplitude was reduced when a large perpendicular magnetic field was applied (Fig. 3c) because magnetic field suppresses superfluid density and hence Ic in the junction. However, owing to the much smaller area of the nano-bridge comparing with tri-layer-based junctions, such suppression was mitigated in our nano-SQUIDs, which would be functional as a magnetometer under 5 KGs of external field. Under zero magnetic field, the Flux noise of our nano-SQUIDs at 4.2K with a SQUID array amplifier operating under FFL was about $1\mu\Phi_0/\sqrt{Hz}$ (Fig. 3d), which was similar to the white noise floor of a tri-layer-based nano-SQUID under similar conditions[26].

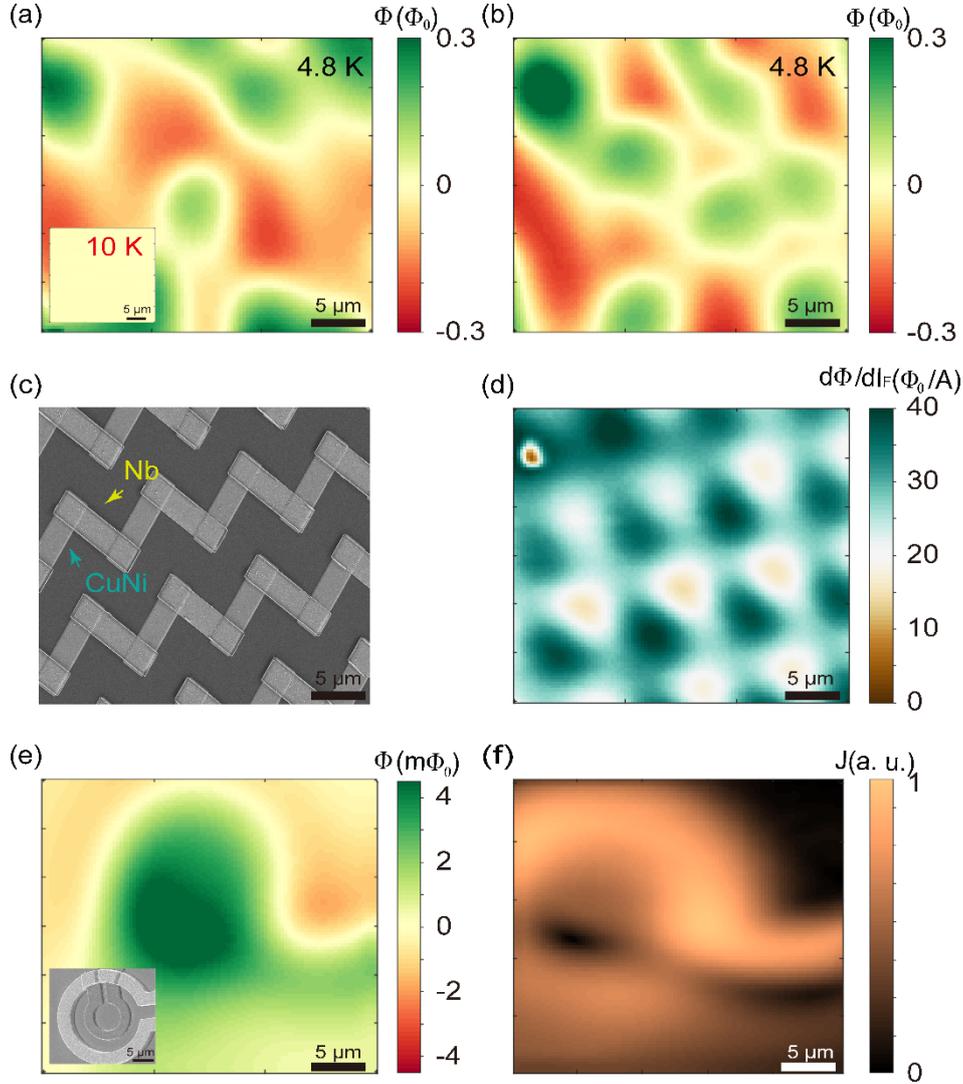

**Fig. 4. Scanning SQUID magnetometry images.** Flux images of a Nb film at 4.8 K (a), after warming up to 10 K (a, inset) and cooling down again to 4.8 K (b). (c) and (d) are the optical and susceptibility image, respectively, of a zigzag test pattern made up of Nb and CuNi bars, which showed up as diamagnetic and ferromagnetic sections in susceptometry. (d) showing alternate ferromagnetism and diamagnetism. (e) Current flux image of the tip of a nano-SQUID when 15 μA current was passed in the field coil. (f) Current density reconstructed from (e) by an FFT algorithm (see text)[27].

We integrate our nano-SQUIDs into a closed-cycle He-4 cryostat for scanning SQUID microscopy. We hand-polish nano-SQUIDs and glue it to a cantilever that acts a capacitive sensor for height. The capacitance is approximately 1pF when the cantilever is relaxed and changes when the SQUID touches the sample. We use a balanced capacitance bridge[28] to detect such a small change in capacitance. The cryostat is shielded by mu-metal to reduce the magnetic field on the sample and the SQUID.

The SQUID is shunted by a small resistance and its current is inductively coupled to a SQUID array amplifier. The output of the array amplifier goes into room temperature feedback electronics which controls the current of modulation coils for a FFL detection of the flux signal. The susceptometry is performed by passing a low frequency current ($I_F$) through the field coil and by demodulating the susceptibility ($d\Phi/dI_F$) out of the flux signal. The current magnetometry can be carried out by passing an AC current ($I_{AC}$) through the sample and by demodulating the current flux signal ($\Phi_I'/I_{AC}$) from the flux signal similarly.

We tested the performance of our scanning SQUID microscope for magnetometry, susceptometry and current magnetometry imaging. For magnetometry, we used a Nb film cooled under zero applied magnetic field (Fig. 4a). We observed strong magnetic contrast when the sample was at 4.8 K. Such contrast disappeared when the sample was warmed up to 10 K (Fig. 4a inset) and reappeared in a different configuration when the sample was cooled back down to 4.8 K (Fig. 4b). Such magnetic contrast from the cooling cycle suggests the presence of trapped vortices and we estimated from the density of these vortices that the remnant field on the sample was about 0.6 Gs. We carried out the susceptometry on a CuNi/Nb zigzag grid patterned by optical lithography (Fig. 4c). The periodic paramagnetic and diamagnetic regions from the CuNi and Nb bars could be well resolved from the susceptibility image (Fig. 4d). We used a nano-SQUID as the sample to test current magnetometry (Fig. 4e inset). We passed $I_{AC}$ = 15 μA through the front field coil and measured the current flux around the tip of the nano-SQUID (Fig. 4e). We reconstructed the current density distribution by performing fast-Fourier transform (FFT) algorithm of the current flux image (Fig. 4e) using the point-spread function of the pickup loop as the kernel. The similarity of the current density image with the SEM image of the field coil (Fig. 4e inset) suggested the validity of our measurement and algorithm.

## Conclusion and outlook

In conclusion, we have designed and fabricated a type of nano-SQUID susceptometers based on Nb 3D-nanobridge junctions. Our fabrication process was simpler than that of the nano-SQUIDs based on tri-layer junctions; whereas the flux noise was on the same level. Our field measurement suggests that these SQUID could potentially function as magnetometers under 5 KGs of perpendicular magnetic field. We used these nano-SQUIDs for scanning SQUID microscopy and demonstrated their capabilities for magnetometry, susceptometry and current magnetometry imaging, which are important for determining the electro-magnetic properties of quantum materials.

By combining deep-etching with EBL defined pickup coils, we might be able to fabricate nano-SQUIDs with better than 1 μm spatial resolution. Our fabrication process can be readily transferred to niobium nitride[29] technologies to extend the working temperature and magnetic field range of future nano-SQUIDs.

## Acknowledgement

Work done at SIMIT was supported by the Frontier Science Key Programs of CAS (Grant No. QYZDY-SSW-JSC033), the National Key R&D Program of China (2017YFF0206105), and the Strategic Priority Research program of CAS (Grant No. XDA18000000). Work done at Fudan University was sponsored by the Ministry of Science and Technology of China (2016YFA0301002 and 2017YFA0303000) and National Science Foundation of China (11827805).